\documentclass[prb,aps,superscriptaddress,twocolumn]{revtex4-2}

\usepackage{amsmath}
\usepackage{amsfonts}
\usepackage{amssymb}
\usepackage{txfonts}
\usepackage{bm}
\usepackage{tabularx}
\usepackage{graphicx,color}
\usepackage{ulem}
\usepackage{mathrsfs}

\usepackage{multirow}
\usepackage{here}

\begin{document}

\title{Rotational Gr\"{u}neisen ratio: A probe for quantum criticality in anisotropic systems}

\author{Shohei Yuasa}
\thanks{These authors contributed equally to this work.}
\author{Yohei Kono}
\thanks{These authors contributed equally to this work.}
\author{Yuta Ozaki}
    \affiliation{
        Department of Physics, Faculty of Science and Engineering, Chuo University, Bunkyo, Tokyo 112-8551, Japan
    }

\author{Minoru Yamashita}
    \affiliation{
        The Institute for Solid State Physics, The University of Tokyo, Kashiwa, Chiba 277-8581, Japan
    }

\author{Yasuyuki Shimura}
\author{Toshiro Takabatake}
    \affiliation{
        Department of Quantum Matter, Graduate School of Advanced Science and Engineering,\\ Hiroshima University, Higashi-Hiroshima 739-8530, Japan
    }
\author{Shunichiro Kittaka}
\thanks{kittaka@g.ecc.u-tokyo.ac.jp}
    \affiliation{
        Department of Physics, Faculty of Science and Engineering, Chuo University, Bunkyo, Tokyo 112-8551, Japan
    }
    \affiliation{
        Department of Basic Science, The University of Tokyo, Meguro, Tokyo 153-8902, Japan
    }

\date{\today}

\begin{abstract}
The Gr\"{u}neisen ratio $\Gamma$ and its magnetic analog, the magnetic Gr\"{u}neisen ratio $\Gamma_H$, are powerful probes 
to study the nature of quantum phase transitions.
Here, we propose a Gr\"{u}neisen parameter, the rotational Gr\"{u}neisen ratio $\Gamma_\phi$, 
by introducing the orientation of the external field as a control parameter.
We investigate $\Gamma_\phi$ of the highly anisotropic paramagnets CeRhSn and CeIrSn by measuring the rotational magnetocaloric effect in a wide range of temperatures and magnetic fields. 
We find that the $\Gamma_\phi$ data of both compounds are scaled by using the same critical exponents and the field-invariant critical field angle. 
Remarkably, the scaling function for the $\Gamma_\phi$ data reveals the presence of highly anisotropic quantum criticality 
that develops as a function of the easy-axis component of the magnetic field from the quantum critical line.
This paper provides a thermodynamic approach to detect and identify magnetic quantum criticality in highly anisotropic systems.
\end{abstract}

\maketitle

\section{INTRODUCTION}

Quantum criticality arising from quantum fluctuations has been a key concept for understanding exotic condensed matter systems such as heavy fermion systems, quantum spin liquids, and high-$T_{\rm c}$ superconductors~\cite{Lohneysen2007RMP,Stewart2001RMP,Gegenwart2008NatPhys,Gegenwart2016RPP,Gegenwart2017PM,Vojta2018RPP,Marel2003Nature}. 
The universal scaling of various quantities in a wide range of temperatures plays an important role in characterizing quantum critical points (QCPs) 
where the system undergoes a quantum phase transition at absolute zero by tuning a control parameter such as pressure $p$, doping concentration, or magnetic field $H$.
In particular, whereas the specific heat $C$, the thermal expansion $\alpha$, and the temperature $T$ derivative of magnetization $M$ approach zero according to the third law of thermodynamics, 
the Gr\"{u}neisen ratio $\Gamma = \alpha/C$ and the \textit{magnetic} Gr\"{u}neisen ratio $\Gamma_H=-(\partial M/\partial T)/C$ have been theoretically predicted to diverge with a universal temperature dependence, proportional to $T^{-1/(\nu z)}$, toward a QCP~\cite{Zhu2003PRL}. 
Here, the correlation length exponent $\nu$ and the dynamical exponent $z$ characterize the divergence of the correlation length $\xi\sim |r|^{-\nu}$ and the correlation time $\xi_\tau\sim\xi^z$ as a function of the control parameter $r$ describing the distance from the QCP, respectively. 
As studied in various materials~\cite{Gegenwart2016RPP,Gegenwart2017PM},
precise estimations of these critical exponents are essential because these exponents reflect the nature of quantum fluctuations arising near the QCP. 
Because of the thermodynamic relation of $\Gamma=[1/(V_m T)](\partial T/\partial p)_S$ and $\Gamma_H=(1/T)(\partial T/\partial H)_S$, 
where $V_m$ is the molar volume, $\Gamma$ ($\Gamma_H$) is related to the piezocaloric effect [magnetocaloric effect (MCE)] and 
becomes a sensitive probe to detect a QCP in which the pressure (magnetic field) is the control parameter. 

In contrast to these previous QCP studies where the {\it magnitude} of the external parameters is relevant, 
the {\it direction} of the external parameters (such as magnetic field, pressure, and electric field) can become an important parameter for the quantum critical behavior when the system has a large anisotropy. 
An introduction of a large anisotropy in the system often gives rise to significant changes in the physical properties that emerge.
For example, a strong local easy-axis anisotropy in the pyrochlore structure yields Ising-like spin correlations, which has been theoretically predicted to realize a quantum spin-ice state with emergent photons~\cite{Gingras2014RPP}.
The bond-dependent Ising anisotropy in the honeycomb structure realizes a Kitaev spin Hamiltonian with fractionalized spin excitations~\cite{Trebst2022PhysRep}.
Therefore, quantum criticality in systems with a strong magnetic anisotropy may exhibit exotic quantum critical phenomena not observed in isotropic systems.
Indeed, the reentrant superconductivity observed in the highly anisotropic magnetic compounds URhGe~\cite{Levy2005Science,Levy2007NatPhys} and UTe$_2$~\cite{Ran2019NatPhys, Tokiwa2024PRB} has been pointed out to be driven by anisotropic spin fluctuations.
However, quantum criticality in these anisotropic systems has not yet been well studied experimentally, due to the lack of appropriate probes.

In this paper, we introduce the \textit{rotational} Gr\"{u}neisen ratio
\begin{equation}
\Gamma_\phi \equiv \frac{1}{T}\biggl(\frac{\partial T}{\partial \phi}\biggl)_S \label{eq:RMCE}
\end{equation}
to study the quantum critical behaviors in which the angle $\phi$ of the external parameters is the control parameter. 
The measurements of $\Gamma_\phi$ have the advantage of minimizing misalignment effects in a highly anisotropic material by varying the angle over a wide range. 
Furthermore, when $\phi$ is the angle of the external magnetic field, adiabatic rotational MCE $({\partial T}/{\partial \phi})_S$ is easier to achieve than adiabatic MCE $({\partial T}/{\partial H})_S$ 
because, unlike the limitation on the ramp rate of the magnetic field by the large inductance of the superconducting magnet, the ramp rate of the rotation angle can easily be set large to realize adiabatic rotational MCE measurements.
We note that $\Gamma_\phi$ given by the magnetic field angle corresponds to the ratio between the temperature derivative of the magnetic torque (${\bm \tau} = -{\bm M} \times {\bm H}$) and $C$ as $(\partial \tau/\partial T)/C$, 
in a similar form of $\Gamma_H = -(\partial M/\partial T)/C$.

We apply this experimental method to the highly anisotropic paramagnets CeRhSn~\cite{Kim2003PRB,Tou2004PRB,Schenck2004JPSJ,Tokiwa2015SA,Kuchler2017PRB,Kittaka2021JPSJ} and CeIrSn~\cite{Tsuda2018PRB,Shimura2021PRL}. 
In CeRhSn, a zero-field QCP is suggested to be realized due to the geometrical frustration caused by the quasikagome lattice~\cite{Tokiwa2015SA}. 
We determine $\Gamma_\phi$ through the measurements of the rotational MCE $(\partial T/\partial \phi)_S$ by rotating the magnetic field in the $ac$ plane. 
We reveal that the $\Gamma_\phi$ data of these two materials follow the same universal relation with the critical exponents $\nu z=2/5$ 
and the same critical field angle $\phi_{\rm cr}=\pi/2$ despite the very different Kondo temperatures (240~K for CeRhSn and 480~K for CeIrSn),
showing a common origin of the quantum criticality in the quasikagome Kondo-lattice system which has antiferromagnetic spin correlation~\cite{Tou2004PRB,Shimura2021PRL} and Ising-type anisotropy ($M_{\parallel c} > M_{\parallel a}$)~\cite{Kim2003PRB,Tsuda2018PRB}. 
We further find that 
the scaling function for the $\Gamma_\phi$ data shows the presence of a \textit{quantum critical line} for both compounds, 
in which the magnetic field along the easy-magnetization axis is the dominant control parameter of the highly anisotropic quantum criticality.
These results demonstrate that $\Gamma_\phi$ is a powerful probe to study quantum criticality in materials with a large magnetic anisotropy.

\section{EXPERIMENTAL Methods}
Single-crystalline samples of CeIrSn and CeRhSn were grown by the Czochralski method~\cite{Kim2003PRB,Tsuda2018PRB}.
A single crystal of CeIrSn with a mass of 14.4 mg weight was used in this study.
The chemical stoichiometry of the single crystals, as well as the negligible sample dependence, was confirmed by the electron-probe microanalysis in a previous study~\cite{Tsuda2018PRB}.
The single crystalline sample of CeRhSn is identical to the one used in the previous study~\cite{Kittaka2021JPSJ}.
The specific heat was measured by the quasi-adiabatic method using a home-built calorimeter installed in a dilution refrigerator (Kelvinox AST minisorb, Oxford).
The rotational MCE was estimated at each field angle $\phi$ by fitting the initial slope of the temperature change, $dT/d\phi$, 
in response to the quasi-adiabatic rotation of the external magnetic field in the $ac$ plane by $1.0^\circ$ or $0.5^\circ$ at a rate of $0.04^\circ-0.1^\circ$/s \cite{Kittaka2018JPSJ}.
Here, $\phi$ is measured from the $c$ axis.
The magnetization of CeIrSn was measured using a capacitive Faraday-force magnetometer~\cite{Sakakibara1994JJAP,Shimizu2021RSI} installed in a $\mathrm{^3}$He refrigerator (HelioxVL, Oxford). 

\section{RESULTS and DISCUSSION}

\begin{figure}
\includegraphics[width=3.4in]{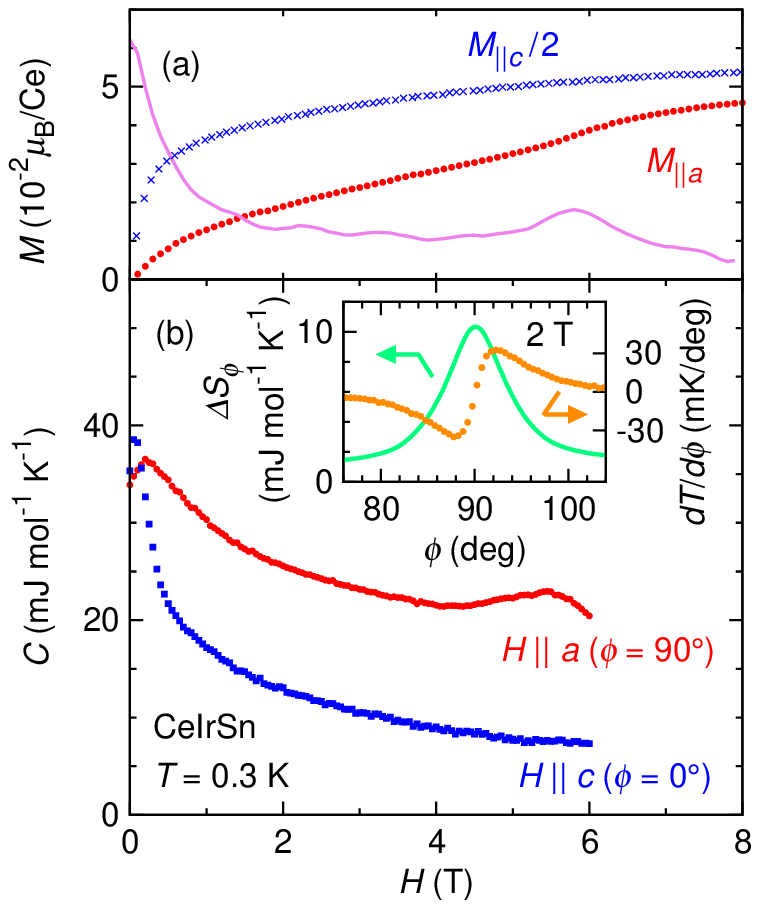}
\caption{
Magnetic field dependence of (a) the magnetization $M$ (circles) for $H \parallel a$ and (b)~the specific heat for $H \parallel c$ ($\phi=0^\circ$) and $H \parallel a$ ($\phi=90^\circ$) of CeIrSn at 0.3~K. 
Here, the field angle $\phi$ is measured from the $c$ axis.
In (a), the field dependence of the derivative of magnetization $dM/dH$ (arb. unit) for $H \parallel a$ is represented by the solid line.
Crosses in (a) show the magnetization $M(H)$ at 0.5~K for $H \parallel c$ taken from Ref.~\onlinecite{Tsuda2018PRB}, which is divided in half.
The inset in~(b) shows the field-angle dependence of the relative entropy change $\Delta S_\phi=S(\phi)-S(0^\circ)$ (solid line) and the rotational MCE $dT/d\phi$ (circles) of CeIrSn under a magnetic field of 2~T rotated within the $ac$ plane at 0.3 K. 
}
\label{fig:M}
\end{figure}

\begin{figure}
\includegraphics[width=3.4in]{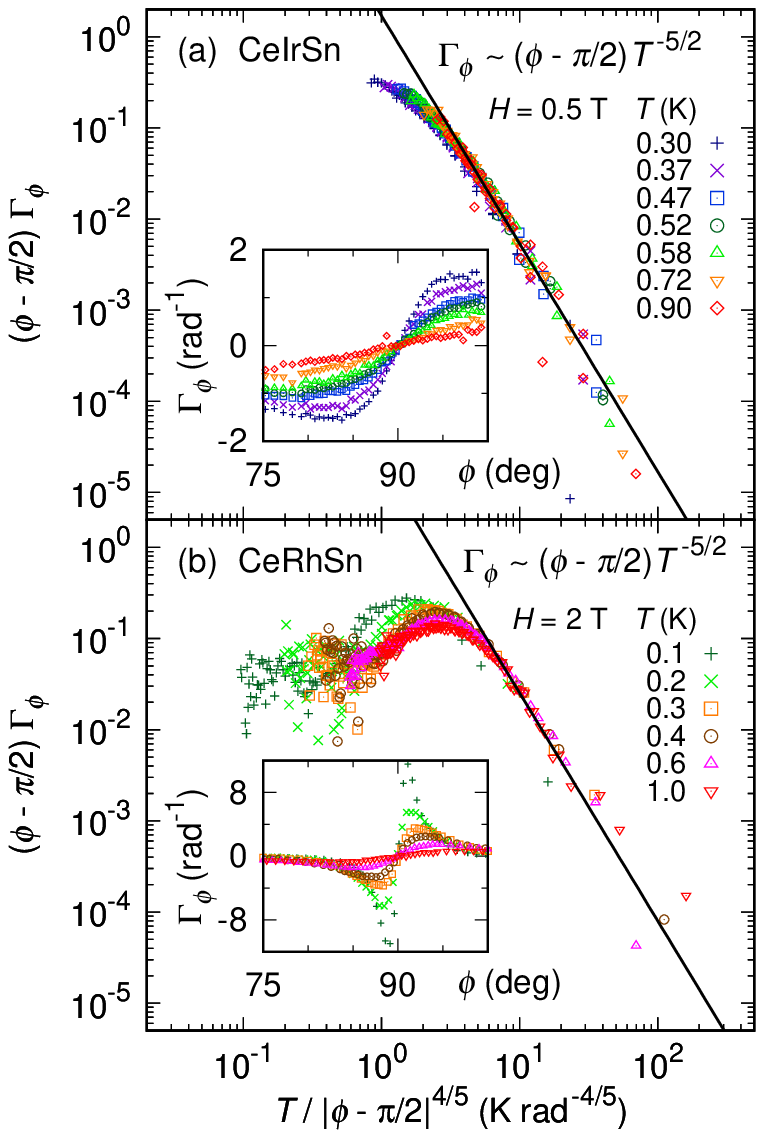}
\caption{
Scaling plots of the rotational Gr\"{u}neisen ratio $\Gamma_\phi$ of (a)~CeIrSn at 0.5 T and (b)~CeRhSn at 2 T for various temperatures.
The solid line indicates the scaling function $\Gamma_\phi \sim (\phi-\pi/2)T^{-5/2}$.
Insets plot the same data of $\Gamma_\phi$ as a function of the magnetic-field angle $\phi$ measured from the $c$ axis.
The rotational MCE data used to estimate $\Gamma_\phi$ are shown in SM (II)~\cite{SM}. 
}
\label{fig:GIr}
\end{figure}

In Fig.~\ref{fig:M}(a), the magnetization of CeIrSn at 0.3~K is plotted as a function of the magnetic field applied approximately along the $a$ axis.
In addition, Fig.~\ref{fig:M}(b) shows the field dependence of the specific heat of the same CeIrSn sample at 0.3~K
under a magnetic field applied precisely along the $c$ or $a$ axis.
A metamagnetic anomaly was clearly detected for $H \parallel a$ around 5.5 (5.8)~T from $C(H)$ ($dM/dH$), in reasonable agreements with the previous report~\cite{Tsuda2018PRB}.
As shown in the inset of Fig.~\ref{fig:M}(b) and Supplemental Material (SM) (I)~\cite{SM}, 
a sharp peak (sign change) was detected at $\phi=90^\circ$ in the field-angle dependent entropy $\Delta S_\phi$ (rotational MCE $dT/d\phi$) for CeIrSn, 
where $\Delta S_\phi$ is estimated by integrating $(\partial S/\partial \phi)_{T,H}=-(C/T)(\partial T/\partial \phi)_{S,H}$.
These sharp angle dependences of $\Delta S_\phi$ and $dT/d\phi$ ensure that the field orientation was precisely controlled during our MCE and specific-heat measurements for CeIrSn (see SM (I)~\cite{SM} for CeRhSn).
The adiabatic condition during the measurements has also been ensured by confirming that the initial slope of $T(\phi)$ just after starting the field rotation was identical 
for the clockwise and the anti-clockwise rotation \cite{Kittaka2018JPSJ, Kittaka2021JPSJ} (see SM (II) for more details).
Given the larger nuclear contribution in the specific heat of CeIrSn at low temperatures than that of CeRhSn~\cite{Tsuda2018PRB}, 
only the data above 0.3~K are presented for CeIrSn in this study.
Using Eq.~\eqref{eq:RMCE} and the rotational MCE data (see SM (II)~\cite{SM} for all the rotational MCE data), the rotational Gr\"{u}neisen ratio $\Gamma_\phi$ of CeIrSn is estimated 
as exemplified in the inset of Fig.~\ref{fig:GIr}(a).
We also estimate $\Gamma_\phi$ of CeRhSn from the rotational MCE data which include those obtained from the previous measurements~\cite{Kittaka2021JPSJ,SM}, 
as exemplified in the inset of Fig.~\ref{fig:GIr}(b).

We now discuss the scaling of the rotational Gr\"{u}neisen ratio $\Gamma_\phi$ for CeIrSn and CeRhSn.
Based on a theoretical model in the vicinity of a QCP~\cite{Zhu2003PRL}, 
if a QCP exists at the critical field angle $\phi_{\rm cr}$ as the applied magnetic field $H$ is rotated with a plane,
a scaling for $(\phi-\phi_{\rm cr})\Gamma_\phi$ is defined by a function of $(\phi-\phi_{\rm cr})\Gamma_\phi \sim f(T/(\phi-\phi_{\rm cr})^{\nu z})$. 
This scaling describes the divergence of $\Gamma_\phi$,
given as $\sim T^{-1/(\nu z)}$ in the quantum critical regime ($T/T_0 \gg r^{\nu z}$) and 
$\sim 1/(\phi-\phi_{\rm cr})$ in the quantum paramagnetic (disordered) regime ($T/T_0 \ll r^{\nu z}$).
Here, the control parameter is $r(\phi)=(\phi-\phi_{\rm cr})/\phi_0$, $T_0$ and $\phi_0$ are constants for scaling, $\phi_{\rm cr}$ is a critical field angle, and $f(x)$ is a scaling function.

We examine this scaling relation for $\Gamma_\phi$ using the data in the insets of Figs.~\ref{fig:GIr}(a) and \ref{fig:GIr}(b).
By investigating the scaling of the $(\phi-\phi_{\rm cr})\Gamma_\phi$ data as a function of $T/(\phi-\phi_{\rm cr})^n$ for various exponents $n$, 
we find that $n=4/5$ and $\phi_{\rm cr}=\pi/2$ provide the best scaling as shown in Figs.~\ref{fig:GIr}(a) and \ref{fig:GIr}(b) for CeIrSn at 0.5~T and CeRhSn at 2~T, respectively.
Thus, $(\phi-\pi/2)\Gamma_\phi$ data at the fixed $H$ collapse on the same curve and $\Gamma_\phi$ shows quantum critical divergence as $(\phi-\pi/2)T^{-5/2}$ for both materials.
In Fig.~\ref{fig:GIr}(b), $(\phi-\pi/2)\Gamma_\phi$ becomes constant in the low $T/(\phi-\pi/2)^{4/5}$ regime, corresponding to the quantum paramagnetic regime.
As shown in the SM (III)~\cite{SM}, the $(\phi-\pi/2)\Gamma_\phi$ data at different magnetic fields also show similar scaling on different curves as a function of $T/(\phi-\pi/2)^{4/5}$. 
These scalings suggest that the correlation length $\xi$ diverges at $\phi_{\rm cr}=\pi/2$ regardless of the magnitude of $H_\perp$ below at least 2~T with the common critical exponents $\nu z$, 
indicating the existence of a quantum critical \textit{line} along $\phi=\pi/2$ (the hard-magnetization $a$ axis).

\begin{figure}
\includegraphics[width=3.4in]{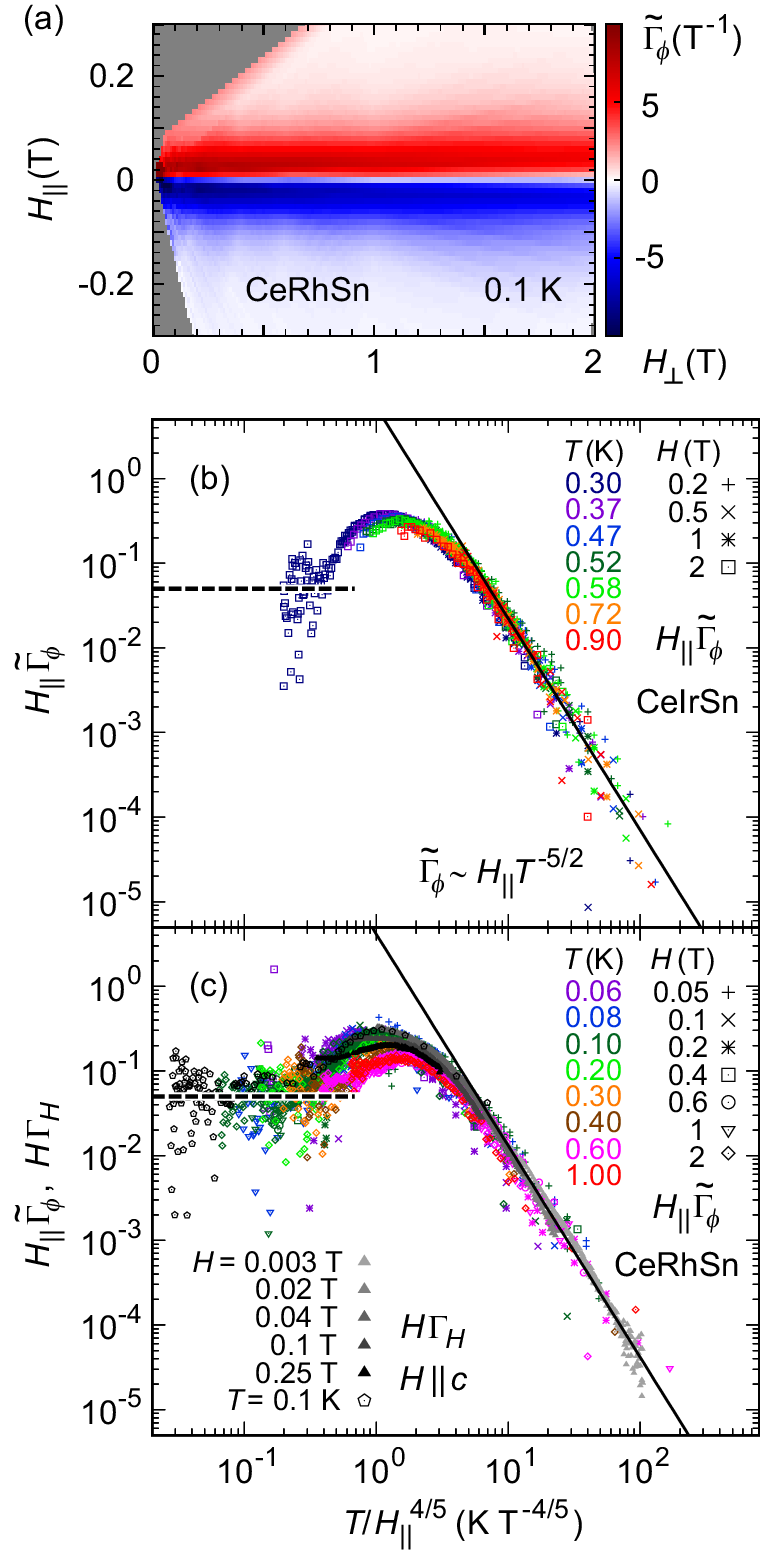}
\caption{
(a) Contour plot of $\tilde{\Gamma}_\phi(H_\parallel,H_\perp)$ for CeRhSn at 0.1~K. The gray region indicates no data available.
(b), (c) 
Scaling plots of the rotational magnetic Gr\"{u}neisen ratio $\tilde{\Gamma}_\phi$ of (b) CeIrSn and (c) CeRhSn for various magnetic fields and temperatures.
The different symbols (colors) of the data correspond to the different magnetic fields (temperatures) of the measured data.
In (c), the magnetic Gr\"{u}neisen ratio $\Gamma_{H}$ of CeRhSn for $H \parallel c$, i.e., $\Gamma_H|_{\phi=0}$, is also plotted. 
The field-dependent $\Gamma_H(H)|_{\phi=0}$ data at 0.1~K (pentagons) were obtained using the MCE data in Ref.~\onlinecite{Kittaka2021JPSJ}, and
the temperature-dependent $\Gamma_H(T)|_{\phi=0}$ data (solid triangles) were taken from Ref.~\onlinecite{Tokiwa2015SA}. 
The solid line indicates the scaling function $\tilde{\Gamma}_\phi \sim H_{\parallel}T^{-5/2}$.
The dashed line represents $H_\parallel\tilde{\Gamma}_\phi=0.05$.
}
\label{fig:Gru}
\end{figure}

This presence of the quantum critical line along $\phi = \pi/2$ indicates that the quantum criticality is dominated by the $c$-axis component of the magnetic field ($H_\parallel = H \cos \phi$), 
but is insensitive to the $a$-axis component ($H_\perp = H \sin \phi$). 
In order to clarify the control parameter in this system, we investigate 
the {\it rotational magnetic} Gr\"{u}neisen ratio $\tilde{\Gamma}_\phi$ given by 
\begin{equation}
\tilde{\Gamma}_\phi \equiv -\frac{\Gamma_\phi}{H_\perp}=-\frac{1}{TH_\perp}\biggl(\frac{\partial T}{\partial \phi}\biggl)_S=\frac{1}{T}\biggl(\frac{\partial T}{\partial H_\parallel}\biggl)_S.
\end{equation}
Therefore, the scaling relation of $\Gamma_H=(1/T)(\partial T/\partial H)_S$ valid at $\phi=0$ can be adopted for $\tilde{\Gamma}_\phi$
in a highly anisotropic system in which the control parameter is given by 
$r(H_\parallel) \sim |H_\parallel - H_{\parallel, {\rm c}}|/H_0$ where $H_{\parallel, {\rm c}}$ is the critical magnetic field along the $c$ axis and $H_0$ is a constant.
As shown in the contour plot of $\tilde{\Gamma}_\phi$ of CeRhSn at 0.1~K in the $H_\parallel$--$H_\perp$ plane [Fig.~\ref{fig:Gru}(a)], 
$\tilde{\Gamma}_\phi$ asymmetrically changes as a function of $H_\parallel$ with respect to $H_\parallel=0$, but it is insensitive to $H_\perp$, 
showing that the magnetic field component along the easy $c$ axis is the dominant control parameter of the quantum criticality.

The scaling of $(H_\parallel -H_{\parallel, {\rm c}})\tilde{\Gamma}_\phi$ as a function of $T/(H_\parallel -H_{\parallel, {\rm c}})^n$ enables to identify the critical field value along the $c$ axis $H_{\parallel, {\rm c}}$.
We find that $n = 4/5$ and $H_{\parallel, {\rm c}}=0$ provide the best scaling for all the data at different magnetic fields as shown in Figs. \ref{fig:Gru}(b) and \ref{fig:Gru}(c) for CeIrSn and CeRhSn, respectively; 
all the experimental $\tilde{\Gamma}_\phi$ data collapse on a universal curve in multiple orders of magnitude of $T/H_\parallel^{4/5}$.
Here, the $H_\parallel\tilde{\Gamma}_\phi$ data for $|\phi|<30^\circ$ are excluded because the error in $\tilde{\Gamma}_\phi$ is enhanced due to the divergence of $H_\perp^{-1}$ around $\phi=0$. 
These scalings with $H_{\parallel, {\rm c}}=0$ again support the presence of the quantum critical line along $H_\parallel=0$ 
(at $\phi_{\rm cr}=\pi/2$) in the $ac$ field plane for both compounds.

For CeRhSn, the magnetic Gr\"{u}neisen ratio $\Gamma_H$ for $H \parallel c$, i.e., $\Gamma_H|_{\phi=0}$, is also plotted in Fig.~\ref{fig:Gru}(c) 
using the reported MCE data~\cite{Tokiwa2015SA,Kittaka2021JPSJ}.
$\Gamma_H|_{\phi=0}$ clearly exhibits the same scaling as $\tilde{\Gamma}_\phi$. 
The difference between $\tilde{\Gamma}_{\phi}$ and $\Gamma_{H}|_{\phi=0}$ is in the contribution from $H_\perp$; 
both $H_\parallel$ and $H_\perp$ vary with the field rotation in the former, 
whereas only $H_\parallel$ is controlled with keeping $H_\perp=0$ in the latter.
Therefore, the common divergence observed both in $\tilde{\Gamma}_{\phi}$ and $\Gamma_{H}|_{\phi=0}$ further supports that the quantum criticality arises mainly from spin fluctuations along the easy-magnetization axis.
Indeed, $\Gamma_H$ for $H \parallel a$ does not follow the present scaling (see SM (IV)~\cite{SM}).

Remarkably, we find the same quantum critical divergence $\tilde{\Gamma}_\phi \sim H_{\parallel}T^{-5/2}$ for both CeIrSn and CeRhSn in the regime $T/H_\parallel^{4/5} \gtrsim 6$~K~T$^{-4/5}$, as represented by the solid lines in Figs.~\ref{fig:Gru}(b) and \ref{fig:Gru}(c).
These results indicate universal quantum criticality emerging in these isostructural materials where a strong Kondo effect and geometric frustration coexist.
In the regime $T/H_\parallel^{4/5}\lesssim 0.3$~K~T$^{-4/5}$, $H_\parallel\tilde{\Gamma}_\phi$ saturates at a constant value, 
as demonstrated by the dashed line in Figs.~\ref{fig:Gru}(b) and \ref{fig:Gru}(c).
This constant $H_\parallel\tilde{\Gamma}_\phi$ shows a formation of a quantum paramagnetic state,
emerging on the universal phase diagram in the vicinity of a QCP~\cite{Vojta2018RPP}.
In this quantum paramagnetic state of CeRhSn, $\tilde{\Gamma}_\phi$ diverges toward $H_\parallel \sim 0$. 
This divergence of $\tilde{\Gamma}_\phi \propto 1/H_\parallel$ supports the presence of a QCP at $H_{\parallel, {\rm c}}=0$, consistent with a zero-field quantum criticality 
suggested by the previous $\Gamma_H$ measurements~\cite{Tokiwa2015SA}.
Because the large nuclear contribution in the specific heat of CeIrSn limited our precise measurements above 0.3 K,
the divergence coefficient of $\tilde{\Gamma}_\phi$ in the low $T/H_\parallel^{4/5}$ region of CeIrSn cannot be well evaluated. 

Let us discuss the critical exponents determined by our measurements done in CeIrSn and CeRhSn.
The universal scaling in the quantum critical regime, $\tilde{\Gamma}_\phi \sim H_\parallel T^{-5/2}$, indicates the exponent $\nu z=2/5$
because of the scaling relation 
$H_\parallel\tilde{\Gamma}_\phi=\tilde{f}(T/H_\parallel^{2\nu z})$ 
expected for a QCP at $H_{\parallel, {\rm c}}=0$~\cite{Tokiwa2014NatMat}.
Here, $\tilde{f}(x)$ is a universal scaling function for $\tilde{\Gamma}_\phi$.
This value of $\nu z$ is also supported by the fact that 
the specific-heat data of CeRhSn for $H \parallel c$ are scaled as a function of $T/H^{2\nu z}$~(see SM (V)~\cite{SM}).

The value of $\nu z=2/5$ obtained for CeRhSn and CeIrSn is much smaller than that expected by the Hertz-Millis-Moriya theory predicting $\nu z = 1$ (3/2) for antiferromagnetism (ferromagnetism)~\cite{Abrahams2014PRB}, as well as $\nu z = 3/4$ found in the heavy-fermion superconductor CeCoIn$_5$~\cite{Tokiwa2013PRL,Tokiwa2014NatMat} and $\nu z = 2/3$ in the quantum spin ice Pr$_2$Ir$_2$O$_7$~\cite{Tokiwa2014NatMat}.
To the best of our knowledge, a theoretical model that consistently explains the critical exponents $\nu z=2/5$ for CeIrSn and CeRhSn is lacking. 
Nonetheless, the small value of the critical exponents $\nu z$ indicates that the correlation length and the correlation time are relatively long. 
These longer correlations may reflect a characteristic of the quantum criticality driven by the geometrical frustration of the quasikagome lattice of these compounds.

Finally, we evaluate $\nu$ and $z$ separately using $\nu z=2/5$ and the scaling relation.
We find the scaling of the specific heat for quantum criticality as $C=T^{d/z}f(T/H_\parallel^{2\nu z})$ with $\nu z=2/5$ and $d/z=0.85$ (Fig.~S8 in SM (V)~\cite{SM}), 
where $d$ is the dimension of the system.
These values lead to $\nu \sim 0.17$ ($0.11$) and $z \sim 2.4$ ($3.5$) for $d=2$~($3$).
We also find $y_0 = 0.8$ (Fig.~S9 in SM (V)~\cite{SM}), from the specific heat data ($C \sim T^{y_0}$) in the quantum disordered regime. 
In addition, the scaling relation of the ac susceptibility $\chi_{\rm ac} \sim H_\parallel^{-\alpha}$ gives $\nu d = 0.35$ 
by adopting $\alpha = 0.5$ (Fig.~S10 in SM (V)~\cite{SM}), where $\alpha=2-2\nu(d+z)$. 
Using these parameters, $G_r = - \nu (d-y_0 z)/y_0$ is estimated to $-0.025$, 
which matches with the experimental $\tilde{\Gamma}_\phi$ data, as demonstrated by a dashed line in Figs. \ref{fig:Gru}(b) and \ref{fig:Gru}(c), where $H_\parallel\tilde{\Gamma}_\phi = -2G_r$. 
Despite the relatively large error, the extracted $G_r$ falls within the expected range, supported by other scaling parameters, 
which provides confidence in the general validity of our analysis.
Moreover, the values of $\nu$ and $z$ will place strong constraints on further theoretical works to clarify the nature of the quantum criticality of these quasikagome compounds.

Before concluding, let us discuss potential developments of this method. 
Compared to the previous measurements done on Pr$_2$Ir$_2$O$_7$ \cite{Tokiwa2014NatMat} and CeCoIn$_5$ \cite{Tokiwa2013PRL}, 
our data exhibits the larger variance due to the lower entropy in CeRhSn and CeIrSn. 
This issue may be resolved by replacing the current monotonic control of the magnetic field angle with an ac modulation, 
as demonstrated in Ref. \onlinecite{Tokiwa2011RSI} for magnetocaloric measurements. 
This approach will be investigated in future work.
This method is expected to be effective in studying finite-field quantum criticality in anisotropic systems, such as Ising magnets, 
and will be explored in future work.

\section{SUMMARY}
We have developed an experimental tool, the rotational Gr\"{u}neisen ratio $\Gamma_\phi$, to characterize quantum criticality sensitive to the external-field angle.
From the rotational magnetocaloric effect, we have investigated $\Gamma_\phi$ of the highly anisotropic quasikagome paramagnets CeRhSn and CeIrSn in a wide range of temperatures and magnetic fields. 
We reveal that the measured $\Gamma_\phi$ data of both compounds are scaled by a function
with the same critical exponents~$\nu z = 2/5$ and $\phi_{\rm cr}=\pi/2$, 
demonstrating the realization of the universal quantum criticality in the geometrically frustrated Kondo-lattice system. 
The field-independent $\nu z$ and $\phi_{\rm cr}$ obtained from the scaling analysis of $\Gamma_\phi$ along with
the good match between the scaling of $\Gamma_H$ at $\phi = 0$ and that of $\tilde{\Gamma}_\phi=-\Gamma_\phi/H_\perp$ suggest
the existence of a quantum critical line in the $ac$ field plane, 
around which the highly anisotropic quantum criticality is dominantly controlled
by the easy-axis component of the magnetic field.
These results indicate that the $\Gamma_\phi$ measurements have the  potential to advance the study of quantum criticality in highly anisotropic systems.

\begin{acknowledgments} 
We thank Y. Tokiwa for useful discussion. The magnetization measurement was carried out as joint research in ISSP. 
This work was supported by KAKENHI (JP23H01128, JP23H04868, and JP24H01673) from JSPS. 
\end{acknowledgments}

\clearpage

\onecolumngrid
\appendix

\begin{center}
{\large Supplemental Material for \\
Rotational Gr\"{u}neisen ratio: A probe for quantum criticality in anisotropic systems}\\
\vspace{0.1in}
Shohei Yuasa,$^1$ Yohei Kono,$^1$ Yuta Ozaki,$^1$ Minoru Yamashita,$^2$\\ Yasuyuki Shimura,$^3$ Toshiro Takabatake,$^3$ and Shunichiro Kittaka$^{1, 4}$\\
{\small 
\textit{$^1$Department of Physics, Faculty of Science and Engineering, Chuo University, Bunkyo, Tokyo 112-8551, Japan}\\
\textit{$^2$The Institute for Solid State Physics, The University of Tokyo, Kashiwa, Chiba 277-8581, Japan \\}
\textit{$^3$Department of Quantum Matter, Graduate School of Advanced Science and Engineering,\\ Hiroshima University, Higashi-Hiroshima 739-8530, Japan}\\
\textit{$^4$Department of Basic Science, The University of Tokyo, Meguro, Tokyo 153-8902, Japan}\\
}
(Dated: \today)
\end{center}

\renewcommand{\thefigure}{S\arabic{figure}}
\renewcommand{\thetable}{S\arabic{table}}
\setcounter{figure}{0}

\subsection*{I. Magnetic-field angle dependence of thermodynamic quantities for CeIrSn and CeRhSn}

\begin{figure}
\includegraphics[width=4.5in]{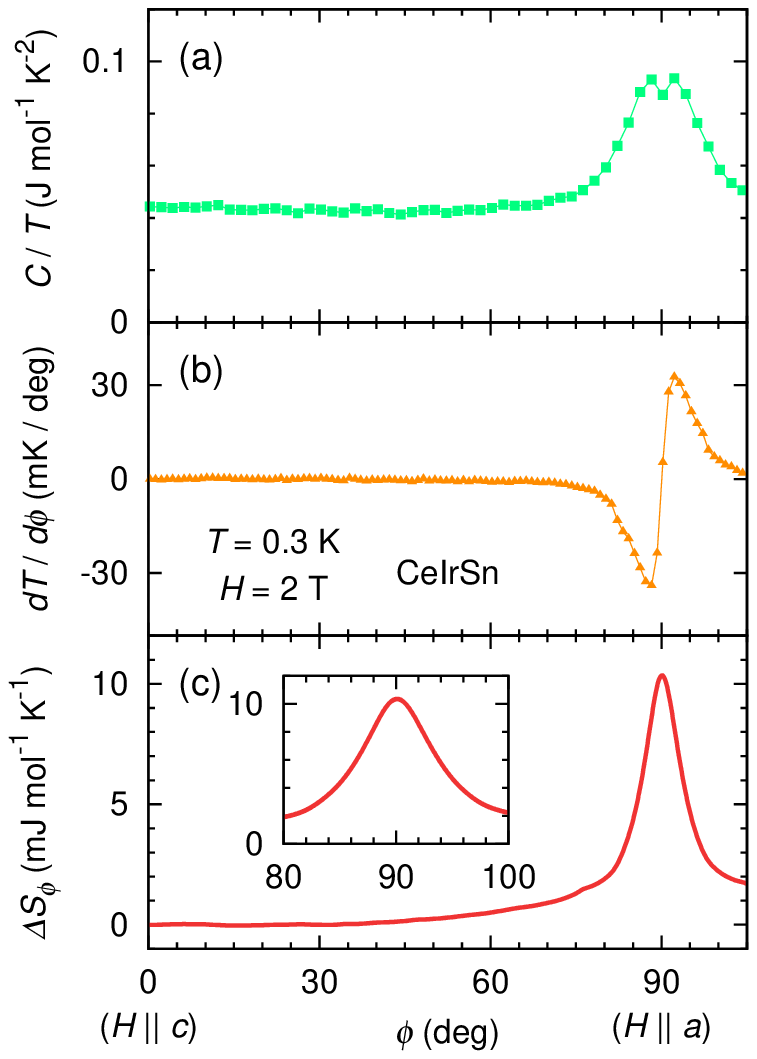}
\caption{
Magnetic-field angle dependence of (a) specific heat divided by temperature $C/T$, (b) rotational magnetocaloric effect $dT/d\phi$, and (c) relative change in entropy $\Delta S_\phi$ of CeIrSn at 0.3~K and 2 T.
The inset in (c) shows an enlarged view of $\Delta S_\phi$.
}
\label{SIr}
\end{figure}

\begin{figure}
\includegraphics[width=4.5in]{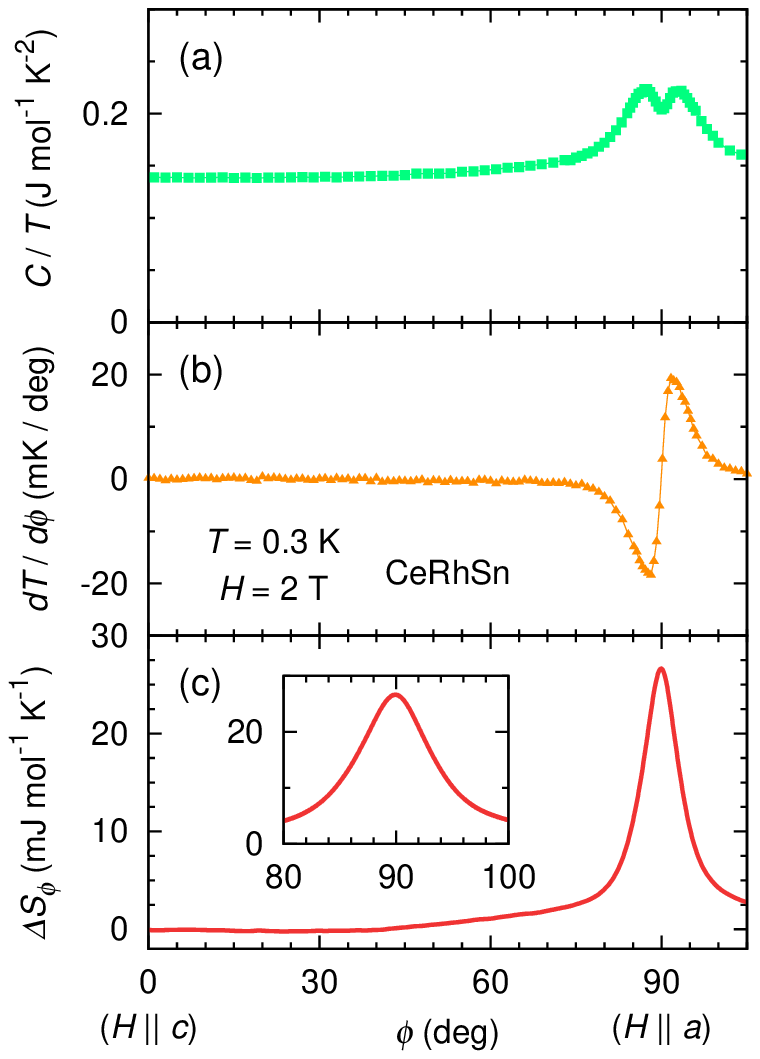}
\caption{
Magnetic-field angle dependence of (a) specific heat divided by temperature $C/T$, (b) rotational magnetocaloric effect $dT/d\phi$, and (c) relative change in entropy $\Delta S_\phi$ of CeRhSn at 0.3~K and 2 T.
The inset in (c) shows an enlarged view of $\Delta S_\phi$.
}
\label{SRh}
\end{figure}

\begin{figure}
\includegraphics[width=4.5in]{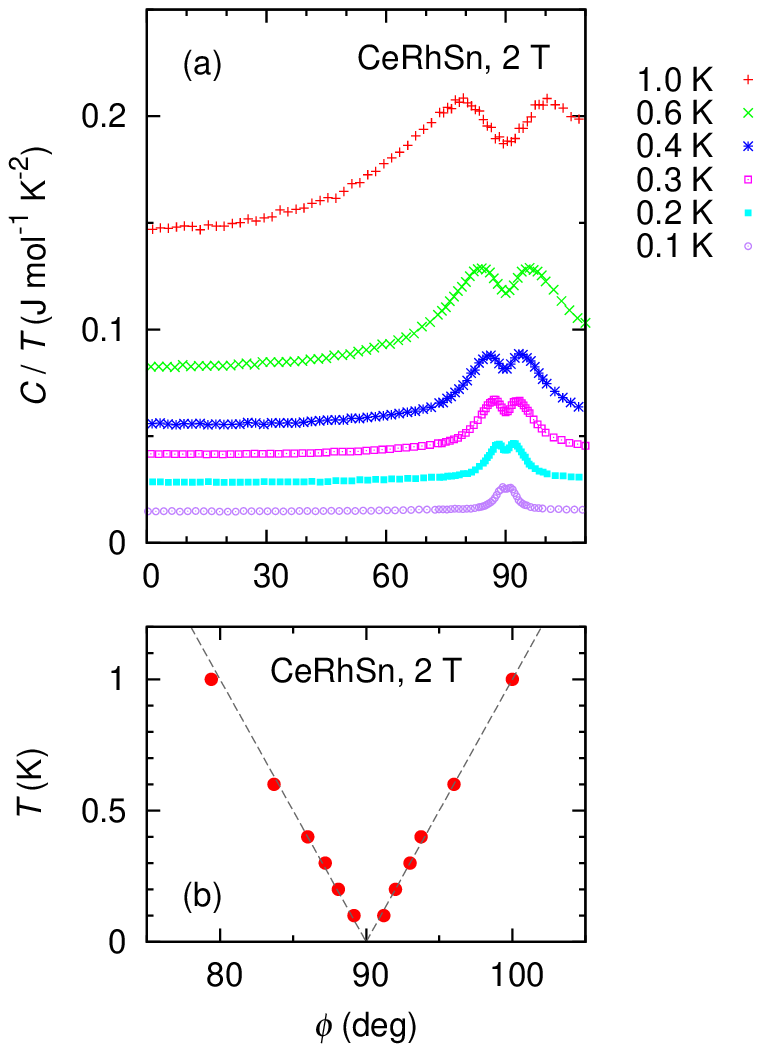}
\caption{
(a) Magnetic-field angle dependence of specific heat divided by temperature $C/T$ at 2 T and various temperatures. (b) The peak positions in $C(\phi)$ at 2 T as a function of temperature.
}
\label{CRhp}
\end{figure}

Figures~\ref{SIr} and \ref{SRh} show the magnetic-field angle dependence of the specific heat, the rotational magnetocaloric effect, and the entropy change measured for CeIrSn and CeRhSn, respectively, at 0.3~K and 2~T, 
indicating qualitatively similar field-angle dependence of all the thermodynamic quantities for both materials.
As shown in Figs. \ref{SIr}(a) and \ref{SRh}(a), the specific heat of CeRhSn is larger than that of CeIrSn at 0.3~K and 2~T.
The anisotropic ratio of the specific heat, $C(90^\circ)/C(0^\circ)$, is 2.0 for CeIrSn, which is lager than 1.5 for CeRhSn.
The magnitude of $dT/d\phi$ of CeIrSn is about 1.8 times larger than that of CeRhSn.
Using the thermodynamic relation, $(\partial S/\partial \phi)_{T,H}=-(C/T)(\partial T/\partial \phi)_{S,H}$,
the relative entropy change, $\Delta S_\phi(\phi)=S(\phi)-S(0^\circ)$, can be evaluated as $\Delta S_\phi=-\int^{\phi}_{0^\circ} (C/T)(dT/d\phi)d\phi$
using the results of specific heat and rotational magnetocaloric effect measurements.
The results are shown in \ref{SIr}(c) and \ref{SRh}(c).
As presented in the inset of Figs.~\ref{SIr}(c) and \ref{SRh}(c), $\Delta S_\phi$ becomes half at $\phi \sim 90^\circ \pm 4^\circ$ for both materials.
The peak seen in $\Delta S_\phi$ becomes sharper on cooling;
e.g., at 0.1 K and 2 T for CeRhSn, $\Delta S_\phi$ becomes half at $\phi=90^\circ \pm 2^\circ$, as reported in the previous study~\cite{Kittaka2021JPSJ}.
As demonstrated in Fig.~\ref{CRhp}, the peak in $C(\phi)$, roughly corresponding to the rapid change in $\Delta S_\phi$, approaches $\phi=90^\circ$ with decreasing temperature to 0~K.

\subsection*{II. Experimental data of the rotational magnetocaloric effect}

\begin{figure}[t]
    \includegraphics[width=6in]{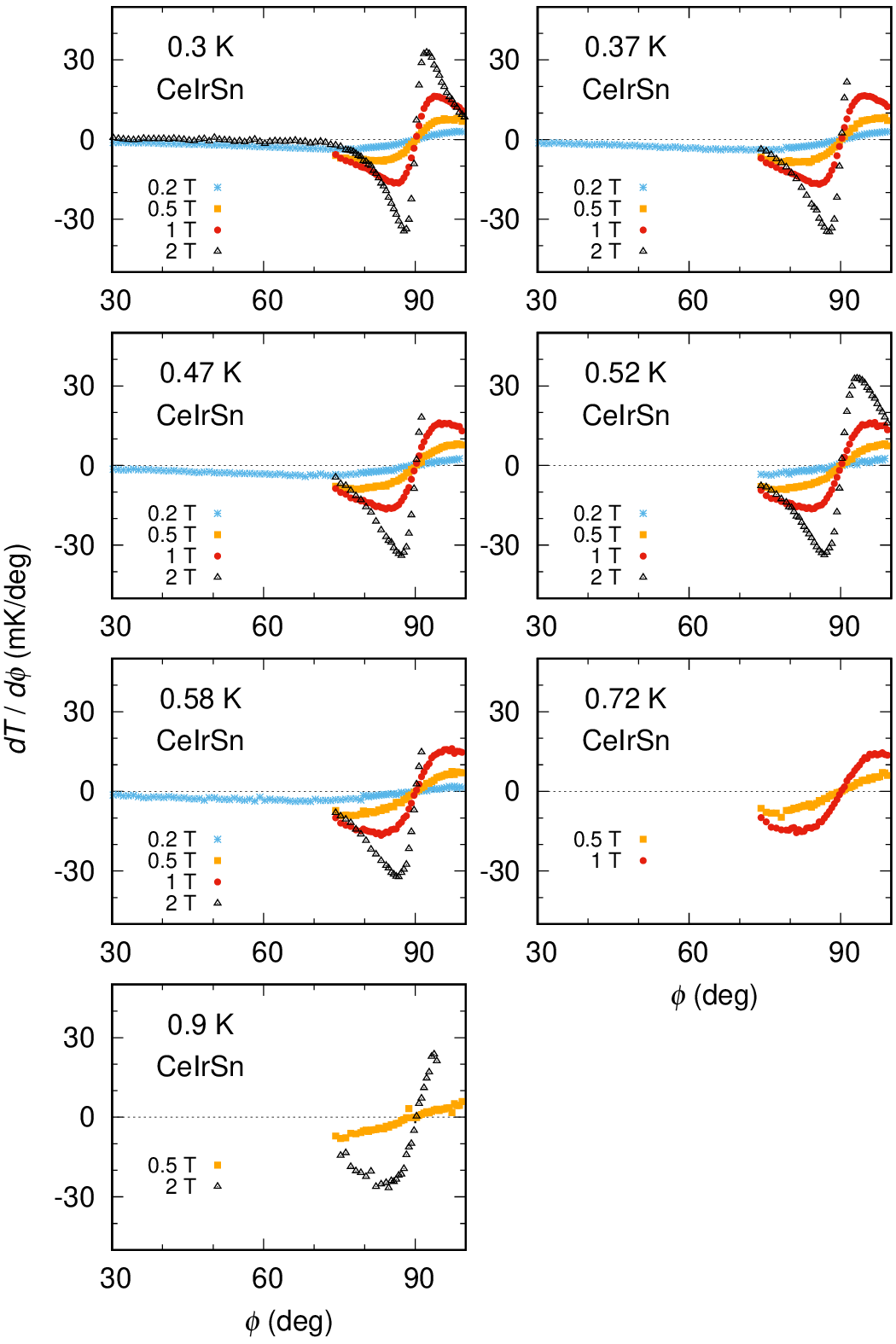}
    \caption{Rotational magnetocaloric effect of CeIrSn at various temperatures and magnetic fields.}
    \label{RMCEI}
\end{figure}

\begin{figure}[t]
    \includegraphics[width=6in]{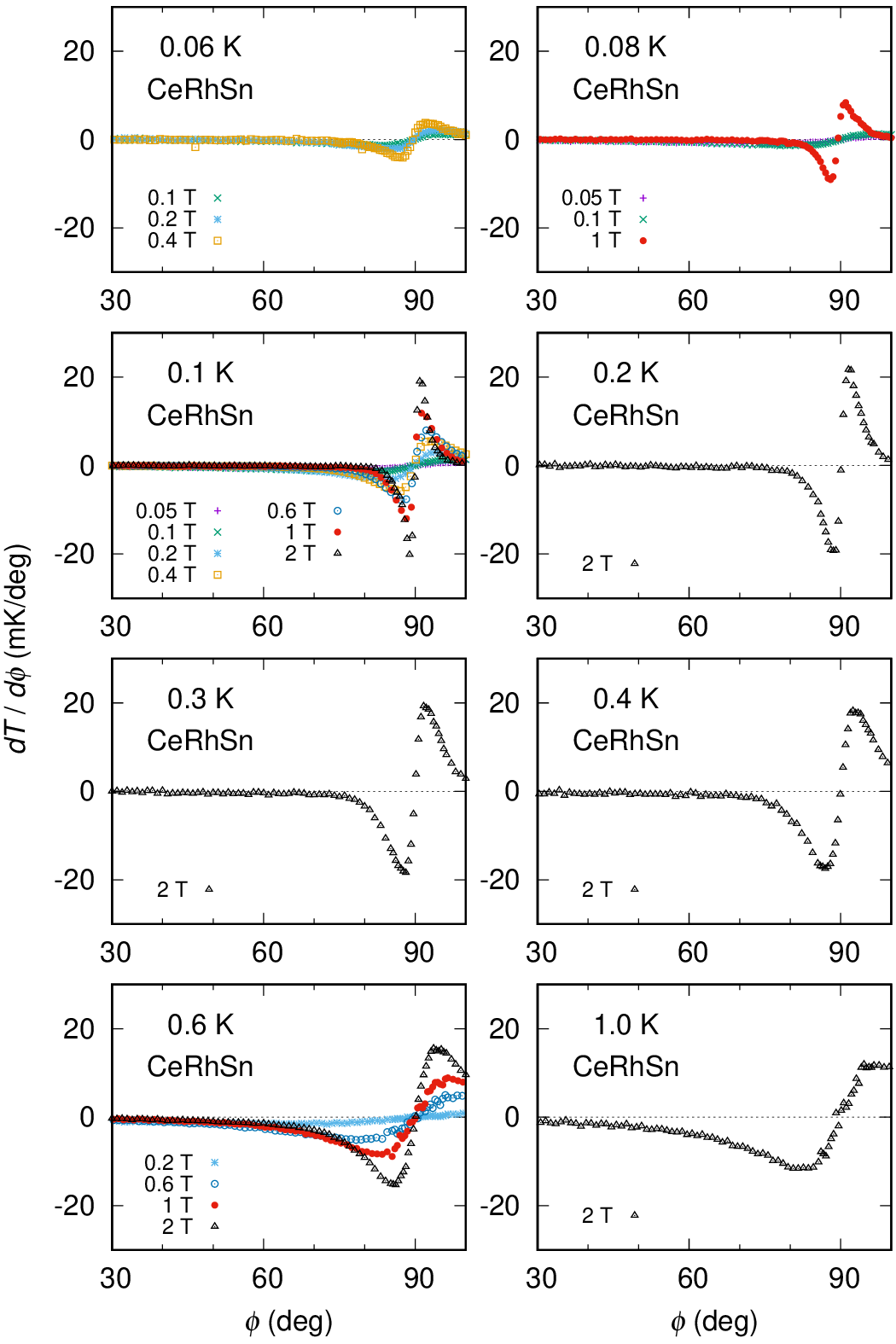}
    \caption{Rotational magnetocaloric effect of CeRhSn at various temperatures and magnetic fields.}
    \label{RMCER}
\end{figure}

Figures~\ref{RMCEI} and \ref{RMCER} display the rotational magnetocaloric effects $dT/d\phi$ of CeIrSn and CeRhSn, respectively, at various temperatures and magnetic fields.
An example of raw data of the temperature change during the magnetic field rotation $T(\phi)$ were reported for CeRhSn in the Supplemental Material (Fig. S2) of Ref.~\onlinecite{Kittaka2021JPSJ}.
The experimental fact that the absolute value of $dT/d\phi$ at $90^\circ+\alpha$ matches well with that at $90^\circ-\alpha$
evidences that the direction of the magnetic field can be accurately controlled. 
This symmetric angle dependence of $dT/d\phi$ also ensures that the adiabatic condition was satisfied 
during our rotational magnetocaloric effect measurements without the heat-transfer effects from the sample or Joule heating by eddy currents.
This is because, as discussed in Ref. \onlinecite{Kittaka2021JPSJ}, whereas the extrinsic heat leakage always increases the temperature, 
the magnetocaloric effect is opposite for rotating clockwise or anticlockwise direction.
In the previous report~\cite{Kittaka2021JPSJ},
the sign of $dT/d\phi$ was apparently opposite because $dT/d\phi$ was incorrectly plotted as a function of $-\phi$, instead of $\phi$ shown as a label of the horizontal axis.
All the $dT/d\phi$ data in the range for $|\phi|>30^\circ$ in Figs.~\ref{RMCEI} and \ref{RMCER} are used to make the scaling plots for the rotational magnetic Gr\"{u}neisen ratio $\tilde{\Gamma}_\phi= -(\partial T/\partial \phi)_S/(H_{\perp}T)$, shown in Fig. 3 of the main text. 

\subsection*{III. Scaling analyses of the rotational Gr\"{u}neisen ratio}

\begin{figure}[t]
    \includegraphics[width=4.5in]{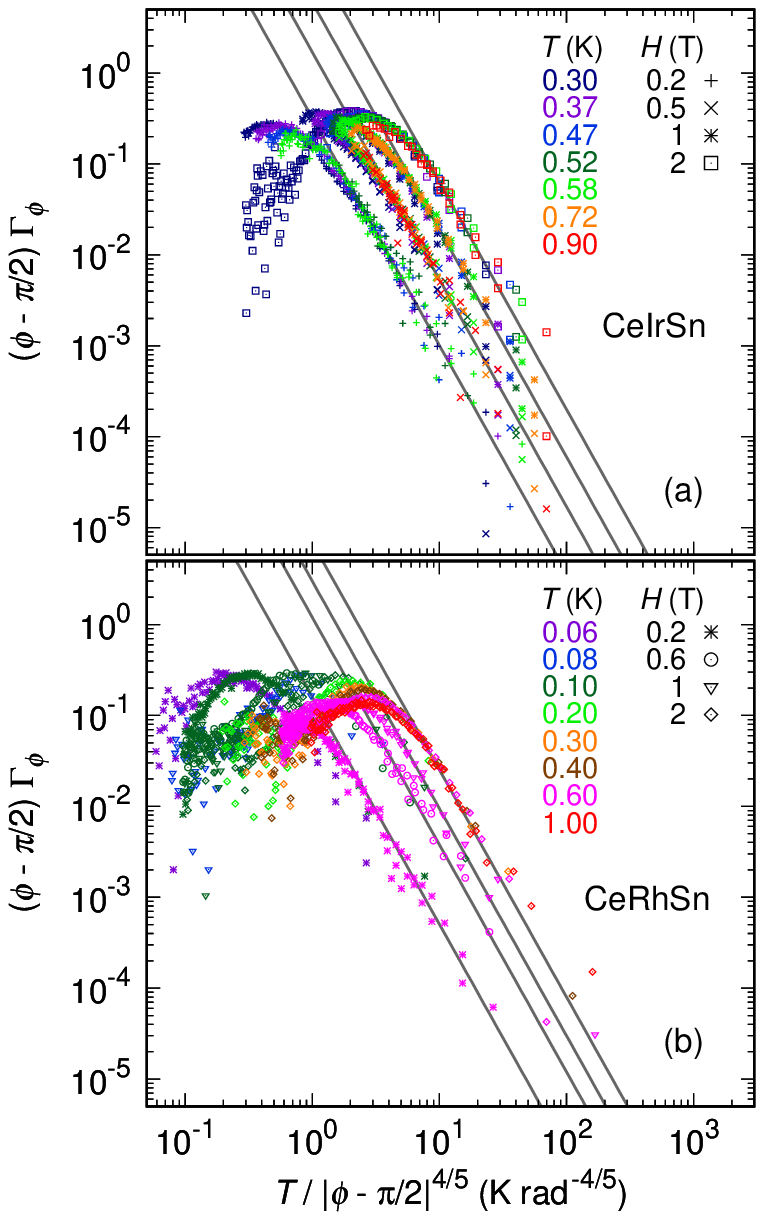}
    \caption{Scaling plots of the rotational Gr\"{u}neisen ratio $\Gamma_\phi=(1/T)(dT/d\phi)$ of (a) CeIrSn and (b) CeRhSn for various temperatures and several selected magnetic fields.
The different symbols (colors) of the data correspond to the different magnetic fields (temperatures) of the measured data.
Solid lines indicate the scaling function $\Gamma_\phi \sim (\phi-\phi_{\rm cr})T^{-5/2}$.
	}
    \label{RGR}
\end{figure}

Figures \ref{RGR}(a) and \ref{RGR}(b) plot $(\phi - \pi/2)\Gamma_\phi$ of CeIrSn and CeRhSn, respectively, as a function of $T/(\phi-\pi/2)^{4/5}$ at various magnetic fields.
The data with the fixed $H$ collapse on the same curve.
It is expected that $\Gamma_\phi(\phi,T)$ data at different magnetic field strengths $H$ collapse on different curves, 
as in the case of $\Gamma_H(H,T)$ when $\phi$ is varied.
In the high $T/(\phi-\pi/2)^{4/5}$ regime, the $(\phi - \pi/2)\Gamma_\phi$ data shows the quantum critical divergence as $\Gamma_\phi \sim A(\phi-\pi/2)T^{-5/2}$.
Here, the coefficient $A$ changes with the magnetic field, as demonstrated by a shift of the solid line in Fig. \ref{RGR}.
Universal scaling of all these data at different magnetic fields is obtained by introducing the rotational magnetic Gr\"{u}neisen ratio $\tilde{\Gamma}_\phi$, as explained in the main text.

\subsection*{IV. Contribution from the perpendicular component of the magnetic field}

\begin{figure}[t]
    \includegraphics[width=4.5in]{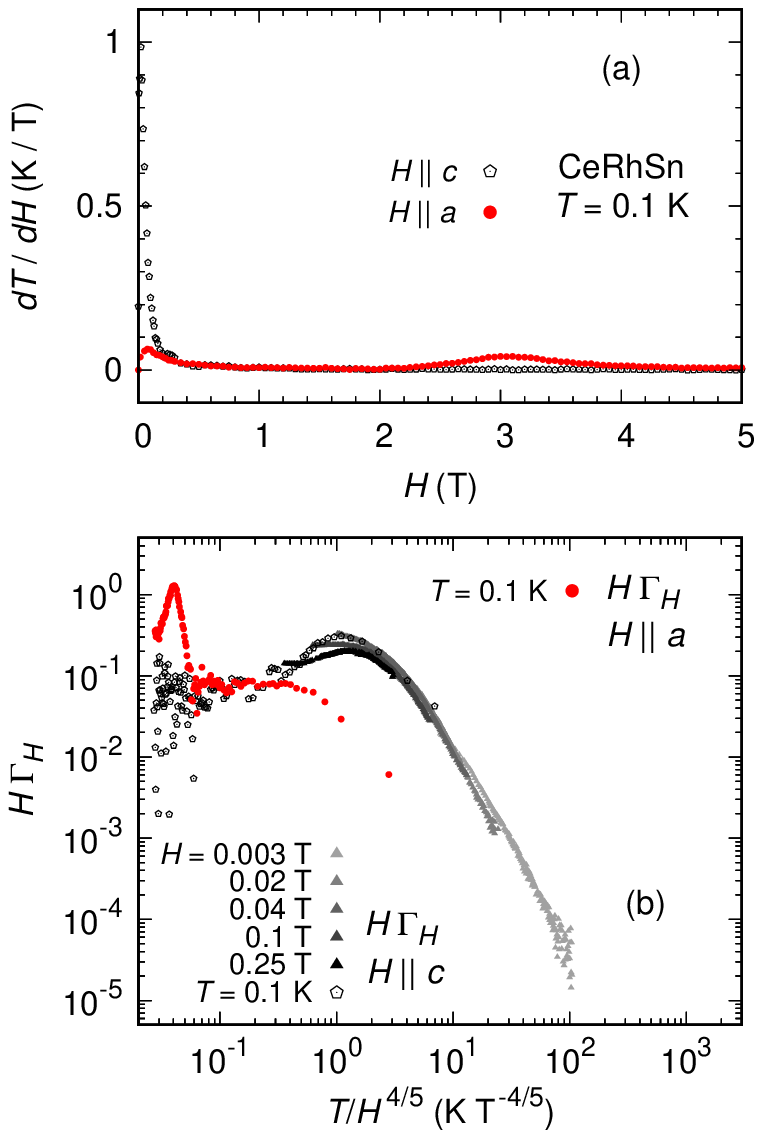}
    \caption{(a) Magnetocaloric effect of CeRhSn at 0.1 K for $H \parallel c$ and $H \parallel a$.
	(b) Scaling plot of the magnetic Gr\"{u}neisen ratio $\Gamma_H$ of CeRhSn for various magnetic fields and temperatures.
	The temperature-dependent data (triangles) were taken from Ref.~\onlinecite{Tokiwa2015SA}. Circles (pentagons) represent the field-dependent $\Gamma_H$ data for $H \parallel a$ ($H \parallel c$).
	}
    \label{MCE}
\end{figure}

Figure \ref{MCE}(a) shows the magnetocaloric effect $dT/dH$ taken at 0.1 K under magnetic fields along the $c$ and $a$ axes~\cite{Kittaka2021JPSJ}.
A small hump observed in $dT/dH$ around 3~T only for $H \parallel a$ can be attributed to the metamagnetic crossover.
Figure \ref{MCE}(b) presents the scaling plot of the magnetic Gr\"{u}neisen ratio $\Gamma_H$ evaluated using the magnetocaloric effect in Fig.~\ref{MCE}(a) and the temperature-dependent magnetocaloric effect for $H \parallel c$ reported previously~\cite{Tokiwa2015SA}.
Clearly, $\Gamma_H$ for $H \parallel a$ does not follow the scaling relation suggested for the $\tilde{\Gamma}_\phi$ and $\Gamma_H|_{\phi=0}$ data in the main text.

The behavior of $dT/dH$ for $H \parallel a$ in Fig.~\ref{MCE}(a) is qualitatively different from the previous report~\cite{Tokiwa2015SA}.
This discrepancy may originate from a slight difference in the field orientation.
Even a slight misalignment can have a significant effect, particularly near $\phi=90^\circ$ ($H \parallel a$), as demonstrated in Fig.~\ref{SRh}. 

\subsection*{V. Evaluation of the critical exponents based on specific heat and ac susceptibility measurements}

On theoretical grounds, the critical contribution to the free energy $F_{\rm cr}$ is expressed as 
\begin{equation}
F_{\mathrm{cr}}= -\rho_0 \biggl(\frac{T}{T_0}\biggl)^{(d + z)/z} f\biggl(\frac{r}{(T/T_0)^{1/(\nu z)}}\biggl)\ = -\rho_0 r^{\nu(d+z)}\tilde{f}\biggl(\frac{T}{T_0r^{\nu z}}\biggl), \label{eq:energy}
\end{equation}
where $\rho_0$ and $T_0$ are constant, $f(x)$ and $\tilde{f}(x)$ are the universal scaling functions, and $d$ is the dimension of the system \cite{Zhu2003PRL}.
Here, let us consider the case of zero-field quantum criticality whose control parameter can be approximated as $r(H) =cH^2$ with $H_{\rm c}=0$ in the limit of small $H$ \cite{Tokiwa2014NatMat}, 
instead of the generic form $r(H)=(H-H_{\rm c})/H_0$, 
where $c$ is a constant coefficient.
Then, the universal scaling behavior can be derived as
$\Gamma_H \sim H T^{-1/(\nu z)}$
in the quantum critical regime for $T/T_0 \gg |r|^{\nu z}$ where $f(x) \approx f(0)+ xf^{\prime}(0)+\cdots$ and
$\Gamma_H = -2G_r/H$
in the quantum paramagnetic regime for $T/T_0 \ll |r|^{\nu z}$ where $\tilde{f}(x) \approx \tilde{f}(0)+ax^{y_0+1}$ ($a$ is constant).
Here, $y_0$ corresponds to the exponent of the power-law behavior of $C(T) \sim T^{y_0}$ when $T/T_0 \ll |r|^{\nu z}$, and $G_r$ is given by $G_r=-\nu(d-y_0z)/y_0$.
As a consequence, $H \Gamma_H$ plotted as a function of $T/H^{2\nu z}$ can be scaled on a universal curve.

The Gr\"{u}neisen ratio in the quantum critical regime includes only one scaling parameter which is the product of $\nu$ and $z$
because the other scaling parameters disappear by taking the ratio of two physical properties. 
Therefore, the critical exponents $\nu z$ can be more easily determined with higher reliability by using the Gr\"{u}neisen ratio.
This is one of the remarkable advantages of studying QCPs using the Gr\"{u}neisen ratio. 
In addition, the coefficient of the magnetic Gr\"{u}neisen ratio in the quantum paramagnetic regime allows us to determine $G_r=-\nu(d-y_0z)/y_0$. 
One can further determine each value of $\nu$ and $z$ by combining the universal scaling relations of other physical quantities. 
It is also important to examine the critical exponents from other scaling relations to improve the reliability of the determinations of these critical exponents.

\begin{figure}[t]
    \includegraphics[width=4.5in]{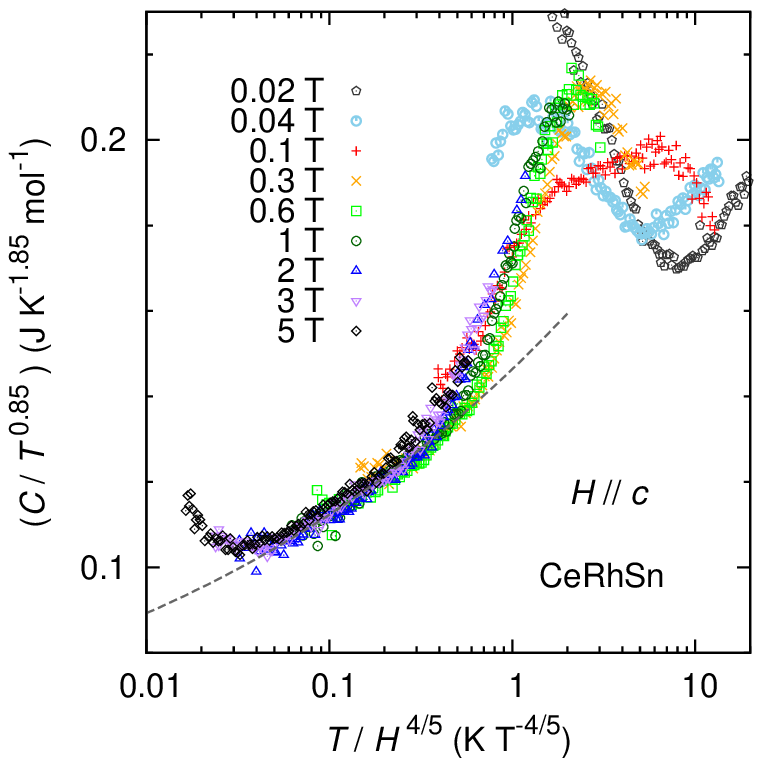}
    \caption{
Scaling plot of the specific heat $C$ of CeRhSn for $H \parallel c$.
The dashed line represents the function $C(T)=bT^{0.8}+\gamma T$.
}
    \label{scaleC}
\end{figure}

In order to further examine the critical exponents, we have performed scaling analyses of the specific heat $C$ of CeRhSn for $H \parallel c$ \cite{Kittaka2021JPSJ}.
On the basis of eq.~\eqref{eq:energy}, the scaling function is expected to behave as
\begin{equation}
C(T,r=0)=\frac{(d+z)d}{z^2}\frac{\rho_0}{T_0}f(0)\biggl(\frac{T}{T_0}\biggl)^{d/z} \sim T^{d/z} 
\end{equation}
for $T/T_0 \gg |r|^{\nu z}$ and 
\begin{equation}
C(T\rightarrow0,r)=\frac{\rho_0ay_0(y_0+1)}{T_0}r^{\nu(d-y_0z)}\biggl(\frac{T}{T_0}\biggl)^{y_0} \sim \biggl(\frac{T}{H^{2\nu z}}\biggl)^{y_0-d/z}T^{d/z}
\end{equation}
for $T/T_0 \ll |r|^{\nu z}$.
Thus, we obtain the universal scaling relation
\begin{equation}
\frac{C(T,H)}{T^{d/z}}=g(T/H^{2\nu z})
\end{equation}
for zero-field quantum criticality, where $g(x)$ is a scaling function. 
As shown in Fig. \ref{scaleC}, we find that $\nu z=2/5$ and $d/z=0.85$ provide a reasonable scaling particularly in the low $T/H^{4/5}$ regime,
although the constant behavior of $C(T,H)/T^{d/z}$ in the quantum critical regime ($T/T_0 \gg |H|^{2\nu z}$) was not observed probably due to noncritical contribution.
Thus, the critical exponents obtained from the specific-heat scaling are consistent with those from the scaling analyses of the rotational Gr\"{u}neisen ratio.
The obtained values of $\nu z=2/5$ and $d/z=0.85$ lead to the critical exponents $\nu \sim 0.17$ ($0.11$) and $z \sim 2.4$ ($3.5$) for $d=2$~($3$).

\begin{figure}[t]
    \includegraphics[width=4.5in]{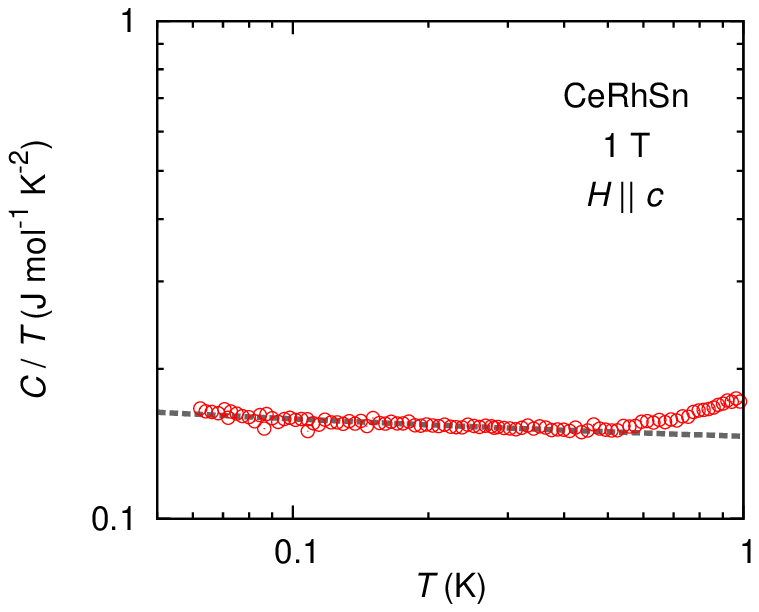}
    \caption{
	Temperature dependence of the specific heat divided by temperature, $C/T$, of CeRhSn at 1 T for $H \parallel c$. 
The dashed line represents the function $C(T)/T=bT^{-0.2}+\gamma$.
}
    \label{fitC}
\end{figure}

\begin{figure}[t]
    \includegraphics[width=4.5in]{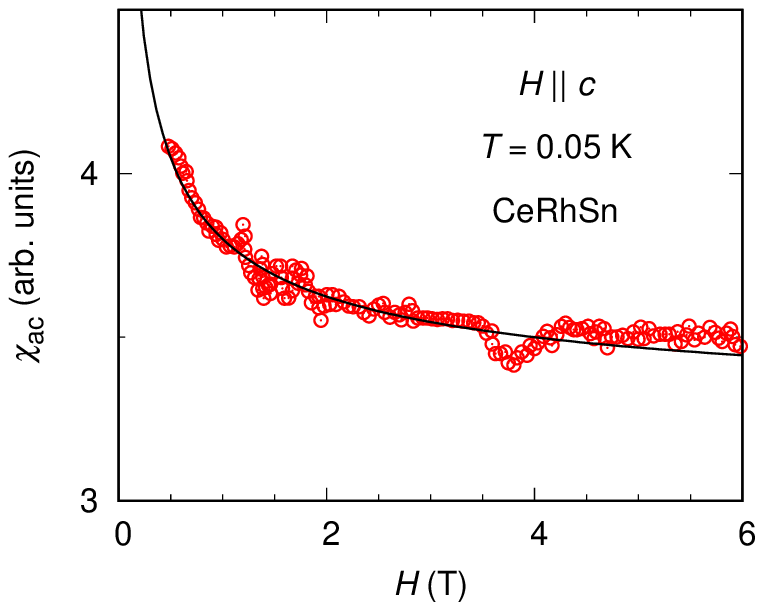}
    \caption{
	Magnetic field dependence of the ac susceptibility $\chi_{\rm ac}$ of CeRhSn at 0.05 K for $H \parallel c$. 
	The data were taken from Ref.~\onlinecite{Yang2017PRB}. 
	The solid line is the function $\chi_{\rm ac}(H)=a_0 H^{-0.5} + b_0$, where $a_0$ and $b_0$ are constants.
}
    \label{chiac}
\end{figure}

Figure~\ref{fitC} displays the temperature dependence of the specific-heat data $C/T$ of CeRhSn at 1~T for $H \parallel c$.
We find that the data can be fitted using the function $C(T) = bT^{y_0}+\gamma T$ with $y_0=0.8$ in the low-temperature range, 
where $b$ and $\gamma$ are constants.
This function also fits the data at different magnetic fields, as shown by the dashed line in Fig.~\ref{scaleC}.
When $\nu z=2/5$, this exponent $y_0=0.8$ leads to $\nu d=0.34$ from the relation $G_r = - \nu (d-y_0 z)/y_0 \sim -0.025$ (see the main text).
This result matches with $d/z=0.85$ obtained above.
Furthermore, in the low-temperature limit, the ac susceptibility $\chi_{\rm ac}$ under a magnetic field is expected to behave as $\chi_{\rm ac} = \partial^2F_{\rm cr}/\partial H^2 \sim H^{-\alpha}$, 
where the scaling relation $\alpha=2-2\nu(d+z)$ should be satisfied.
The obtain exponent $\alpha \sim 0.5$ from the relation is also compatible with the field dependence of $\chi_{\rm ac}$ of CeRhSn for $H \parallel c$ at 0.05~K in Fig.~\ref{chiac}~\cite{Yang2017PRB}.
These facts support the critical exponents $\nu z=2/5$ described in the main text and 
further strengthen the validity of the scaling analyses of the rotational Gr\"{u}neisen ratio, which have been performed for the first time in the present study.

\clearpage

\clearpage

\end{document}